\documentclass[referee]{aa} 

\usepackage{graphics}
\usepackage{epsfig}
\usepackage{natbib}

\begin{document}


\title{The electronic spectra of protonated PANH molecules}


\author{J.~A. Noble \and C. Dedonder \and C. Jouvet}
\institute{CNRS, Aix-Marseille Universit\'{e}, UMR-7345, Physique des Interactions Ioniques et Mol{\'e}culaires (PIIM): 13397 Marseille Cedex 20, France. christophe.jouvet@univ-amu.fr}

\date{Received \today / Accepted \today}

\abstract 
{} 
{This study was designed to examine the viability of protonated nitrogen-substituted polycyclic aromatic hydrocarbons (H$^+$PANHs) as candidates for the carriers of the diffuse interstellar bands (DIBs).} 
{We obtained the electronic spectra of two protonated PANH cations, protonated acridine and phenanthridine, using parent ion photo-fragment spectroscopy and generated theoretical electronic spectra using \textit{ab initio} calculations.} 
{We show that the spectra of the two species studied here do not correspond to known DIBs. However, based on the general properties derived from the spectra of these small protonated nitrogen-substituted PAHs, we propose that larger H$^+$PANH cations represent good candidates for DIB carriers due to the expected positions of their electronic transitions in the UV-visible and their narrow spectral bands.} 
{} 

\keywords{astrochemistry - Molecular data - Ultraviolet: ISM - ISM: molecules - Methods: laboratory: molecular} 

\maketitle

\section{Introduction}\label{sec:intro}

Polycyclic aromatic hydrocarbons (PAHs) are believed to be widely distributed throughout the interstellar medium (ISM) in both the gas and solid phases \citep{Tielens08,Salama08}. The observation of two sets of unidentified spectral features, the diffuse interstellar bands (DIBs) and the unidentified infrared bands (UIRs), is routinely attributed to the presence of PAHs (in absorption and emission, respectively). 

The DIBs -- weak unidentified absorption features in the near ultraviolet (UV), visible, and near infrared (IR) region of the electromagnetic spectrum -- were first observed in the 1920s \citep{Heger22,Merrill34}. These bands are observed on lines of sight containing sufficiently high column densities, such as those traversing diffuse interstellar clouds.
Additionally, the UIR emission features at 3.3, 6.2, 7.7, 8.6, and 11.3~$\mu$m, first observed in the 1970s \citep[e.g.][]{Gillett73}, are observed in the IR spectra of almost all objects. Although no individual carrier molecule has been identified with certainty, the group of polyaromatic hydrocarbons
(PAHs) is believed to contribute to these bands \citep[e.g.][and references therein]{Allamandola89,Salama96}. The development of the so-called ``PAH hypothesis'', which holds PAHs responsible for the UIRs \citep{Leger84} due to UV-pumped IR fluorescence, was very quickly followed by an hypothesis suggesting that ionised PAHs were likely carriers of the DIBs \citep{Leger85,Allamandola85}. 

Both experimental and theoretical studies support the identification of PAHs, and more specifically, PAH cations, as the carriers of the DIBs and UIRs \citep[e.g.][]{Pino99,Salama99,Mattioda05,Pathak08,Garkusha11,Gredel11}. Models have shown that UV pumping \citep[e.g.][]{Schutte93}, or even NIR absorption (in areas of low UV flux \citet{Mattioda08}), can drive MIR emission from PAHs, effectively linking DIBs with UIRs. One of the remaining points of debate is the nature of the PAH cations: whether they are radical cations (PAH$^+$) or protonated species (H$^+$PAH). Radical cations have been shown experimentally to be highly reactive and thus will react readily with hydrogen to form protonated species \citep{Snow98}; PAHs are, however, stripped of H atoms by interaction with photons \citep{Tielens08}.

More recent IR studies have suggested that the sub-group of nitrogen-substituted PAHs, or PANHs, exhibit spectral features similar to those of PAHs, and may also contribute to unidentified spectral bands \citep{Hudgins05, Bernstein05, Mattioda08, Alvaro10}. One study in particular illustrates the importance of PANHs in photon-dominated regions (PDRs), as the spectra of PANH cations were not merely useful, but required, to fit the 6.2 and 11~$\mu$m emission features observed towards NGC~7023 \citep{Boersma13}. This is not an unexpected result, as the abundance of N relative to C in the ISM is $\sim$~0.25 \citep{Spitzer78} and thus PANHs are likely relatively abundant interstellar species.
Additionally, \citet{Hudgins05} conclude that the substitution of N in PAHs could account for the variations observed in the peak position of the 6.2~$\mu$m UIR. 
These spectroscopic studies in the IR have strengthened the case for the presence of charged PANHs in the ISM and their contribution to the UIRs, but very few studies have been made of the electronic spectra of gas phase PANH cations for comparison with the DIBs \citep{Dryza12}.

PANHs have a high proton affinity, so it is highly likely that protonated PANHs exist in the ISM, particularly in ionised environments. It has been illustrated that, for a small sample of PANH molecules, the IR spectra of protonated PANH cations (H$^+$PANH) better reproduce the 6.2 and 8.6~$\mu$m UIRs than the spectra of ionised PANH radical cations (PANH$^+$) \citep{Alvaro10}. However, to date no studies have determined the electronic spectra of H$^+$PANH for comparison with the DIBs.

In this study, we provide the electronic spectra of gas phase protonated cations acridine ([C$_{13}$H$_{9}$N]H$^+$, hereafter AcH$^+$) and phenanthridine ([C$_{13}$H$_{9}$N]H$^+$, hereafter PhH$^+$), nitrogen-substituted versions of the PAHs anthracene and phenanthrene (see Appendix~\ref{numbering}). We show that these protonated molecules present vibrationally resolved electronic states in the visible or near UV spectral region where DIBs are observed.

\section{Methods}

\subsection{Experimental methods}

The electronic spectra of the protonated aromatic PANH were obtained via parent ion photo-fragment spectroscopy in a cryogenically-cooled quadrupole ion trap (Paul Trap from Jordan TOF Products, Inc.) \citep{Alata13}.
 The setup is similar to the one developed in several groups based on the original design by \citet{Wang08}. 

The protonated ions are produced in an electrospray ionisation source built at Aarhus University \citep{Anderson04}. At the exit of the capillary, ions are trapped in an octopole trap for 90~ms. They are extracted by applying a negative pulse of \textit{ca.} 50~V and are further accelerated to 190~V by a second pulsed voltage just after the exit electrode. This timed sequence of pulsed voltages produces ion packets with durations of between 500~ns and 1~$\mu$s. The ions are driven by electrostatic lenses towards the Paul trap biased at 190~V so that the ions enter the trap gently, avoiding fragmentation induced by collisions. A mass gate placed at the entrance of the trap allows selection of the parent ion. The Paul trap is mounted on the cold head of a cryostat (Coolpak Oerlikon) connected to a water-cooled He compressor. Helium gas is injected in the trap using a pulsed valve (General Valve) triggered 1~ms before the ions enter the trap, as previously reported by \citet{Kamrath10}. The ions are trapped and thermalised at a temperature between 20 and 50~K through collisions with the cold He buffer gas. The ions are held in the trap for several tens of ms before the photodissociation laser is triggered. This delay is necessary both to ensure thermalisation of ions and the efficient pumping of the buffer gas from the trap in order to avoid collision induced dissociation of the ions during subsequent extraction. 

The photo-dissociation laser is an optical parametric oscillator (OPO) laser, which has a 10~Hz repetition rate, 10~ns pulse width, a resolution of $\sim$~8~cm$^{-1}$ and a scanning step of 0.02~nm (laser from EKSPLA uab). The laser is shaped to a 1~mm$^2$ spot to fit the entrance hole of the trap and the laser power is around 20~mW in the UV spectral region. 
After laser excitation, the ions are stored in the trap for a delay that can be varied between 20 and 90~ms before extraction into the 1.5~m time-of flight mass spectrometer. The full mass spectrum is recorded on a micro channel plates (MCP) detector with a digitising storage oscilloscope interfaced to a PC. The photofragmentation yield detected on each fragment is normalised to the parent ion signal and the laser power.

\subsection{Calculations}

\textit{Ab initio} calculations were performed with the TURBOMOLE programme package \citep{Ahlrichs89}, making use of the resolution-of-the-identity (RI) approximation for the evaluation of the electron-repulsion integrals \citep{Hattig03}. The equilibrium geometries of the protonated species in their electronic ground states (S$_0$) were determined at the MP2 (M{\o}ller-Plesset second order perturbation theory) level. Adiabatic excitation energies of the lowest electronic excited singlet states (S$_1$) were determined at the RI-ADC(2)(second order Algebraic Diagrammatic Construction level \citep{Schirmer82}). Calculations were performed with the correlation-consistent polarised valence double-zeta (cc-pVDZ) basis set \citep{Woon93}. The vibrations in the ground and excited states were calculated for protonated acridine and phenanthridine and the electronic spectra simulated using the PGOPHER software \citep{Western} for Franck Condon analysis.

\section{Results and discussion}

\subsection{Protonated acridine}

\begin{figure*}[!tb]
\includegraphics[width=\textwidth]{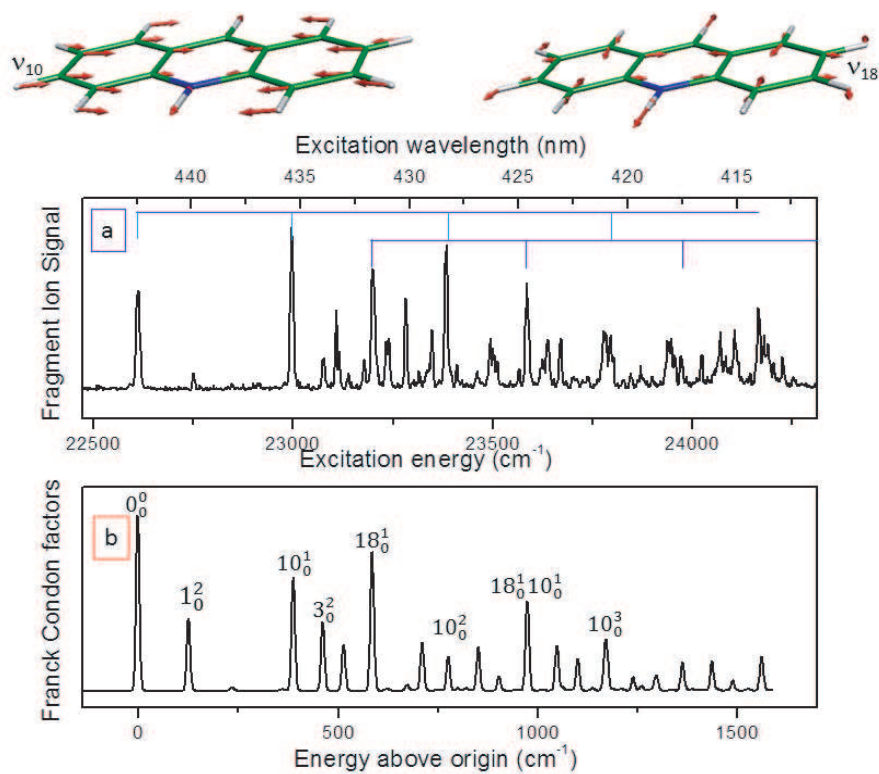}
\caption{Photofragmentation spectrum of protonated acridine compared to the simulated spectrum:  a) the experimental spectrum plotted as the fragment ion signal (arbitrary units) as a function of the excitation energy; b) the simulated spectrum plotted as the Franck-Condon factors as a function of the excitation energy, calculated using the calculated ground and excited state frequencies. The main vibrational progressions (indicated in blue in panel a) involve, essentially, the two symmetric modes $\nu_{10}$ at 383~cm$^{-1}$ (ring breathing) and $\nu_{18}$ at 587~cm$^{-1}$. These vibrational modes are illustrated schematically above the experimental spectrum, with arrows representing the direction and relative magnitude of the displacement of each atom. The intensities of the out of plane modes $\nu_{1}$ and $\nu_{3}$ are overestimated in the simulation.}\label{fig:acridine}
\end{figure*}

The protonated acridine photofragmentation spectrum is presented in Fig.~\ref{fig:acridine}a. The first electronic transition shows clear vibrational progressions in the visible starting at 442.24~nm (22612~cm$^{-1}$/ 2.80~eV) and extending to $\sim$~400~nm, this absorption region being in agreement with the absorption recorded in water solution at low pH \citep{Ryan97}. These progressions involve the 383~cm$^{-1}$ vibrational mode starting from the 0-0 transition ($\nu_0$), and a symmetric mode at 587~cm$^{-1}$ ($\nu_{18}$). The vibronic bands are narrow, with a FWHM of 10~$\pm$~1 cm$^{-1}$ that represents the convolution of the laser band width ($\sim$~8~cm$^{-1}$) with a low temperature rotational contour \textit{i.e.} the apparent line broadening due to the fact that the spacing between rotational lines is narrow compared to the laser width. 

The main fragmentation channels correspond to loss of H, 2H/H$_2$ and m/z~28 (probably corresponding to loss of H$^+$HCN/H$^+$HNC, \citet{Johansson11}) complemented by weaker fragmentation channels: loss of m/z~26, 27, 29 and loss of m/z~51, 52 and 53 (see mass spectra in Appendix~\ref{mass_spectra}).
   
Ground and first excited state calculations were performed, both with DFT/TD-DFT (Density Functional Theory/ Time Dependent-DFT) and MP2/ADC(2) methods, for the isomer protonated on the nitrogen atom, which is by far the most stable. The ground state energies of the isomers protonated on carbon atoms were also calculated using DFT with the B3LYP functional and the cc-pVDZ basis set, and they all lie more than 1.5~eV higher than the isomer protonated on N. The molecule stays planar in both states. The first excited state is calculated vertically at 3.13~eV, in agreement with previous CASPT2 (complete active space perturbation theory) calculations \citep{Rubio01}, and corresponds to a $\pi\pi$* HOMO-LUMO (highest occupied molecular orbital-lowest unoccupied molecular orbital) transition ($^1$L$_a$ in Platt's notation). Optimisation leads to an adiabatic transition at 2.75~eV (2.60~eV when the difference in zero point energy is included), in agreement with the experiment (2.80~eV), which corroborates the location of the proton on the nitrogen atom. 

The changes in geometry between the ground and excited states are minor: for the outer rings, there is an alternation of long/short C-C bonds in S$_0$, which becomes short/long in S$_1$, with a maximum difference in bond length of 0.053~{\AA} for the C1C2 and C7C8 bonds (see Appendix~\ref{numbering} for the S$_0$ and S$_1$ optimised geometries and the numbering of atoms); for the center ring bearing the N atom, the NH bond is unchanged, N10C11 \& N10C14 increase by 0.031~{\AA}, C9C12 \& C9C13 increase by 0.035~{\AA}, while C11C12 \& C13C14 decrease by 0.008~{\AA}. The angles differ by at most 2$^{\circ}$.

The spectrum was simulated (see Fig.~\ref{fig:acridine}b) using the ground and excited state frequencies calculated and the PGOPHER software for Franck Condon analysis \citep{Western}. The agreement between the experimental and simulated spectra is quite good for the band positions, and allows assignment of the main vibrational progressions to the symmetric breathing mode $\nu_{10}$ (contraction/elongation of the rings along the long axis) alone or in combination with another symmetric mode $\nu_{18}$ (elongation of the center ring along the short axis with contraction of the outer rings). The scheme of the vibrations is shown at the top of Fig.~\ref{fig:acridine}. Two out of plane modes are weakly active in the experimental spectrum and overestimated in the simulation: $\nu_{1}$ is the butterfly mode and $\nu_{3}$ mostly an out of plane motion of the N10-H and C9-H atoms (seesawing motion).

\subsection{Protonated phenanthridine}

\begin{figure*}[!tb]
\includegraphics[width=\textwidth]{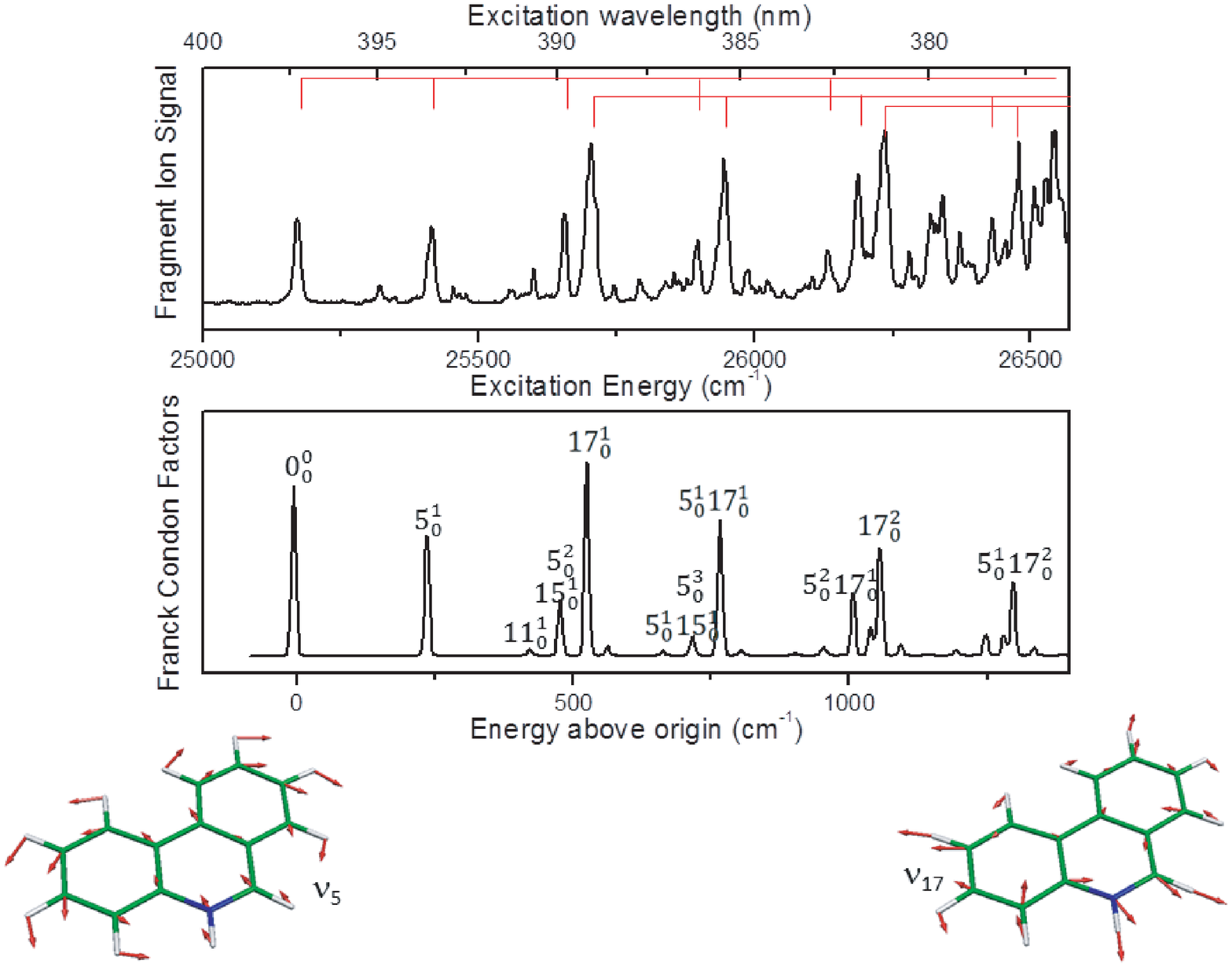}
\caption{Photofragmentation spectrum of protonated phenanthridine compared to the simulated spectrum:  a) the experimental spectrum plotted as the fragment ion signal (arbitrary units) as a function of the excitation energy; b) the simulated spectrum plotted as the Franck-Condon factors as a function of the excitation energy, calculated using the calculated ground and excited state frequencies. The main vibrational progressions (indicated in red in panel a) $\nu_{5}$ and $\nu_{17}$, are illustrated schematically below the simulated spectrum.}\label{fig:phenanthridine}
\end{figure*}

The protonated phenanthridine photofragmentation spectrum (presented in the upper panel of Fig.~\ref{fig:phenanthridine}) shows clear vibrational progressions in the near UV spectral region, with a first transition at 397.28~nm (25171 cm$^{-1}$/ 3.12~eV). This transition is blue shifted by 2500~cm$^{-1}$ compared to the AcH$^+$ transition. Vibrational progressions on a mode of 240~cm$^{-1}$ starting on the 0-0 transition ($\nu_0$), and on a symmetric mode at 533~cm$^{-1}$ ($\nu_{17}$), are observed for several quanta, and the bands are narrow, with a FWHM of 12~$\pm$~1~cm$^{-1}$ (see Appendix~\ref{PhH+vibronic}). As for AcH$^+$, the main fragmentation channels correspond to loss of H, 2H/H$_2$ and H$^+$HCN/H$^+$HNC (see mass spectrum in Appendix~\ref{mass_spectra}).

Ground and first excited state calculations performed for the isomer protonated on the nitrogen atom show that PhH$^+$ stays planar in both states (see Appendix~\ref{numbering}). As for AcH$^+$, this isomer is by far the most stable, the calculated ground state energies of the isomers protonated on carbon atoms (DFT/ B3LYP/ cc-pVDZ) all lie more than 1.6~eV higher than the isomer protonated on N. The first excited state is a $\pi\pi$* state calculated vertically at 3.45~eV and optimisation leads to an adiabatic transition at 3.22~eV, in agreement with the experiment (3.12~ev). 

\subsection{Three-ringed protonated PANHs}

For both protonated acridine and phenanthridine, the proton is located on the nitrogen atom, as shown by the excellent agreement between experiments and calculations: the calculated excited state transitions are within 0.2~eV from the experimental band origins and the spectra simulated with the calculated frequencies reproduce well the experimental progressions. Protonated acridine and phenanthridine are planar in the ground and excited states, which was not the case for protonated anthracene and phenanthrene \citep{Alata10b}. The line widths are narrow (taking into account the laser bandwidth), indicating long lived excited states. 

These two protonated PANHs absorb in the visible or near UV spectral region, this absorption region being in agreement with the absorption recorded in water solution at low pH \citep{Ryan97}. This is the same absorption region as previously recorded for protonated PAHs with two to six aromatic rings, protonated naphthalene \citep{Alata10a}, anthracene \citep{Alata10b,Garkusha11}, phenanthrene \citep{AlataThesis,Garkusha11}, fluorine \citep{Alata12b}, tetracene \citep{Alata10b}, pyrene \citep{AlataThesis,Garkusha11,Hardy13} and coronene \citep{Garkusha11,Rice14}.

\begin{table*}
\caption{Transition origin (eV) for selected neutral and protonated PAHs and PANHs.}\label{table:transitions}
\begin{tabular}{cccc}
\hline\hline
PAH & Transition  & PANH & Transition  \\ 
       & origin (eV)  &           & origin (eV)   \\ 
\hline
Anthracene                & 3.43$^{1}$ & Acridine                     & 3.22$^{2,3}$\\
AnthraceneH$^+$     & 2.52$^{4}$ & AcridineH$^+$           & 2.80$^{5}$\\
\hline
Phenanthrene            & 3.63$^{6}$ & Phenanthridine           & 3.64$^{7}$\\
PhenanthreneH$^+$ & 2.08$^{8,9}$ & PhenanthridineH$^+$ & 3.12$^{5}$\\
\hline
\end{tabular}
\tablebib{
(1) \citet{Lambert84}; (2) \citet{Prochorow98}; (3) \citet{Rubio01}; (4) \citet{Alata10b}; (5) this work; (6) \citet{Hager88}; (7) \citep{Prochorow04}; (8) \citet{AlataThesis}; (9) \citet{Garkusha11}.
}
\end{table*}

Protonation of aromatic heterocycles shifts the absorption spectra to lower energy, but to a smaller extent than for the fully carbonated analogues: for example acridineH$^+$ is red shifted by 0.4~eV from acridine, while anthraceneH$^+$ is red shifted by 0.9~eV from neutral anthracene (see Table~\ref{table:transitions}). This results in a blue shift of the transition of protonated acridine and phenanthridine compared to protonated anthracene and phenanthrene. The reverse was observed in the case of radical cations: ionised quinoline and isoquinoline have their transitions red shifted compared to ionised naphthalene \citep{Dryza12}.

\section{Astrophysical Implications}

\begin{table*}
\caption{Comparison of observed DIB wavelengths and equivalent widths with experimental AcridineH$^+$ wavelengths (in nm).}\label{table:wavelengths}
\begin{tabular}{ccccc}
\hline\hline
\multicolumn{2}{c}{DIB}            & \multicolumn{3}{c}{Experimental AcH$^+$}       \\ 
wavelength & FWHM/{\AA} & wavelength  & intensity\tablefootmark{a}        & Assignment\tablefootmark{b}  \\
\hline
442.883$^{1}$/442.819$^{2}$/442.888$^{3-5}$  & 22.5$^{1,2}$/12.3$^3$ &	442.24 &	100 &	0$_0^0$ \\
                                                                          & &	439.54                                 &	15 &	1$_0^2$ \\
437.173$^{1}$/436.386$^{2}$                            & 1.03/0.46$^{1,2}$ & & & \\
                                                                          & &	434.86                                 &	170 &	10$_0^1$ \\
                                                                          & &	433.38                                 &	30 & \\
                                                                          & &	432.74                                 &	85 &	3$_0^2$ \\
                                                                          & &	432.18                                 &	15 &	\\
                                                                          & &	431.42                                 &	30 &	\\
                                                                          & &	431.04                                 &	130 &	18$_0^1$ \\
                                                                          & &	430.30                                 &	55 &	\\
                                                                          & &	429.52                                 &	95 &	\\
                                                                          & &	428.9                                   &	20 &	\\
                                                                          & &	428.3                                   &	65 &\\	
                                                                          & &	427.62                                 &	150 &	10$_0^2$ \\
                                                                          & &	427.14                                 &	25 &	\\
                                                                          & &	426.22                                 &	20 &	\\
425.901$^{2}$	                                                  & 1.05$^{1,2}$&	425.62                 &	55 &	\\
                                                                          & &	424.32                                 &	20 &	\\
                                                                          & &	423.98                                 &	110 &	10$_0^1$18$_0^1$ \\
                                                                          & &	423.28                                 &	25 &	\\
                                                                          & &	423.02                                 &	50 & \\	
                                                                          & &	422.46                                 &	55 & \\	
                                                                          & &	420.56                                 &	60 &	10$_0^3$ \\
                                                                          & &	420.24                                 &	55 &	\\
                                                                          & &	418.94                                 &	25 &	\\
417.65$^{3-5}$	                                                  & 23.3$^{3}$& 417.62                        &	50 &	10$_0^2$18$_0^1$ \\
                                                                          & &	417.16                                 &	35 &	\\
                                                                          & &	416.24                                 &	35 &	\\
                                                                          & &	415.46                                 &	60 & \\	
                                                                          & &	414.84                                 &	65 & \\	 
                                                                          & &	413.83                                 &	85 &	10$_0^1$18$_0^2$ \\
                                                                          & &	413.5                                   &	45 &	\\
                                                                          & &	412.72                                 &	35 &	\\
                                                                          & &	411.08                                 &	25 &	\\
\hline
\end{tabular}
\tablebib{
(1) \citet{Hobbs09}; (2) \citet{Hobbs08}; (3) \citet{Jenniskens94}; (4) \citet{Krelowski95}; (5) \citet{Tuairisg00}.
}  
\tablefoot{
\tablefoottext{a}{The intensity of the vibronic bands relative to the origin band are given (as percentages), but the absolute values should be used for guidance only, as there are possible saturation effects in the experimental method which are not well controlled.}
\tablefoottext{b}{The notation $v_0^n$ corresponds to a transition from the
 vibrationless level in the ground electronic state, S$_0$, to a level
 with $n$ quanta of the vibrational mode $v$ in the S$_1$ electronic excited
 state. At the low temperature of the experiment, only the
 vibrationless level is populated in the S$_0$ ground state.}}
\end{table*}

In this study we have reported the electronic spectra for two protonated three-ringed PANH cations, derived both experimentally and theoretically. These are the first such experimental spectra, and allow us to draw several astrophysically-relevant conclusions for H$^+$PANH ions. Protonated PAHs have been suggested to be among the potential candidates for diffuse interstellar band (DIB) carriers \citep{Salama96,LePage97,Snow98,Pathak08,Hammonds09} and this study was designed to determine if protonated PANHs could be DIBs carriers by comparing the laboratory gas phase absorptions at low temperature to astronomical data \citep{Hobbs08,Hobbs09}.

The first important conclusion is that H$^+$PANH ions represent good candidates for the source of the DIBs. The substitution of a N atom into the H$^+$PAH skeleton introduces a blue shift in the UV-visible electronic spectrum, resulting in bands at higher energies than those observed for H$^+$PAHs. It has previously been shown that both N-substitution \citep{Hudgins05} and protonation of PAHs \citep{Alvaro10} introduce blue shifts in the IR emission features, giving a better fit to the UIR features attributed to PAHs. A direct comparison of our H$^+$PANH spectra with H$^+$PAH spectra of anthracene-9H and phenanthene-9H \citep{Garkusha11} reveals blueshifts of $\sim$~11 and 133~nm, respectively, due to the presence of N in the ion. The destabilising effect of N on the ion in its excited state is particularly marked for phenanthridine. For the small PANH investigated here, this blue shift results in the majority of their absorption features occurring at higher energies than those observed for DIBs, but this does not preclude the bands of larger, more astrophysically-relevant PANHs falling within the DIBs frequency range. 

Importantly, the absorption lines in our spectra have FWHM of approximately 10~cm$^{-1}$ ($\sim$~15 -- 20~{\AA}), which (upon correction for the $\sim$~8~cm$^{-1}$ bandwidth of the laser) are comparable to observed DIB line widths of 2 -- 3~cm$^{-1}$. Our data are directly comparable to all of the DIBs in the 445 -- 410 nm range, two of which have FWHM of $\sim$~ 23~{\AA} ($>$ 100~cm$^{-1}$) and three of which are on the order of 1~{\AA} (2 -- 5 cm$^{-1}$, see Table~\ref{table:wavelengths}). The narrow bands of these H$^+$PANH species are in direct contrast with those of PAH$^+$ radical cations, which are much wider ($\sim$~20 -- 30~cm$^{-1}$, \citet{Hudgins05}) due to their intrinsically shorter excited state lifetimes. Due to the closed-shell nature of the H$^+$PANH, they are also more photochemically stable than open-shell radical cation species, and therefore would be expected to be more stable in, for example, PDRs. It should be noted here that neutral PAHs (which are themselves closed-shell species) also exhibit narrow absorption lines of $\sim$~3~cm$^{-1}$ \citep{Tan05} and are likely to be major contributors to the weak DIBs.

The first electronic transition that we observe in the spectrum of protonated acridine is centred at 4422.4~{\AA} and has a bandwidth-corrected FWHM of $<$~15~{\AA}. The strongest broad DIB is centred at 4428.2~{\AA} and has a FWHM of 22.50~{\AA} \citep{Hobbs08,Jenniskens94}. The AcH$^+$ band is shifted by $\sim$ 6~{\AA} compared to this unusually broad DIB, and thus we conclude that it is not responsible for the interstellar feature. The other strong bands in the AcH$^+$ spectrum do not match known DIBs either (see Table~\ref{table:wavelengths}).
Moreover, the poor correlation between DIBs precludes the contribution of molecular systems with high vibrational activity to these bands. Although the protonated PANHs studied here present fewer active modes than their fully carbonated analogues, they do show vibrational progressions. The same behaviour seems to exist in small PANH radical cations, which present long vibrational progressions in the visible \citep{Dryza12}. Nonetheless, the progressions that we observe in our small H$^+$PANHs would likely become less pronounced in larger species as they derive from vibrational modes which would be damped by the larger aromatic structure. Additionally, PANHs bend less than the corresponding PAHs, therefore there is more chance of observing isolated bands. Protonation of the studied PANHs occurs preferentially on the N atom, and both species contain N substituted in the PAH exoskeleton. The most favourable protonation orientation was thus H$^+$ in the plane of the molecule, maintaining the planar nature of the species upon ionisation. This would not necessarily be the case for endoskeletal N as \textit{(i)} the presence of N already breaks the planar nature of the molecule and \textit{(ii)} if protonation occurs on the N, it would occur above or below the planar ring structure. The resulting non-planar structure would most likely have additional vibration in electronically excited states, resulting in vibrational progressions in the electronic spectra. Thus, exoskeletal H$^+$PANH probably represent better candidates for the DIBs.

However, as discussed above, it is important to remember that small H$^+$PAH and H$^+$PANH molecules will not be stable in interstellar environments such as PDRs or diffuse HI clouds, and these experiments must be extended to species with larger skeletons. Previous studies suggest that the 6.2~$\mu$m band is best reproduced by PAHs with 60 -- 90 C atoms \citep{Schutte93} and with 2 -- 3 N atoms \citep{Hudgins05}. Such species may be photostable and their electronic transitions should be more to the red, where there is a larger number of DIBs. As mentioned in \S~\ref{sec:intro}, the interstellar abundance of N is $\sim$~0.25 that of C \citep{Spitzer78}. Thus, from a purely statistical point of view, all PAHs large enough to be stable in the ISM (approximately $>$ 50 C) should contain more than $\sim$~ 10 N atoms. Any deviation from these statistical abundances therefore reveals evidence of selectivity in the formation routes to such species. \citet{Hudgins05} conclude that ``most'' PAH species contributing to the UIR features contain nitrogen atoms, with an estimated $>$ 1.2~\% of interstellar N contained in PANHs. When compared to the fraction of elemental C locked up in PAHs (approximately 3.5~\% \citep{Tielens08}), the relative abundance of N appears slightly more than that expected from the statistical N:C ratio, but this could be due to the difference in derivation methods for PAH and PANH abundances.

On the basis of the results of IR \citep{Hudgins05} and UV-visible (this work) studies into PANHs, it is clear that such species are good candidates for inclusion in the ``PAH hypothesis''. Although the spectra of the small H$^+$PANHs studied in this work do not, themselves, correspond to observed DIBs, this preliminary study points to the need for further examination of this class of molecular ion. In particular, the effects of larger PANHs, substitution of larger numbers of N (or other heteroatoms), and endoskeletal heteroatom substitution on the UV-visible and IR spectra of protonated PANH molecules should be investigated.

\begin{acknowledgements}
J.~A. Noble is a Royal Commission for the Exhibition of 1851 Research Fellow. We acknowledge the use of the computing facility cluster GMPCS of the LUMAT federation (FR LUMAT 2764).
\end{acknowledgements}

\clearpage 

\appendix

\section{Structures of acridine and phenanthridine}\label{numbering}

\begin{figure*}[!htb]
\includegraphics[width=0.5\textwidth]{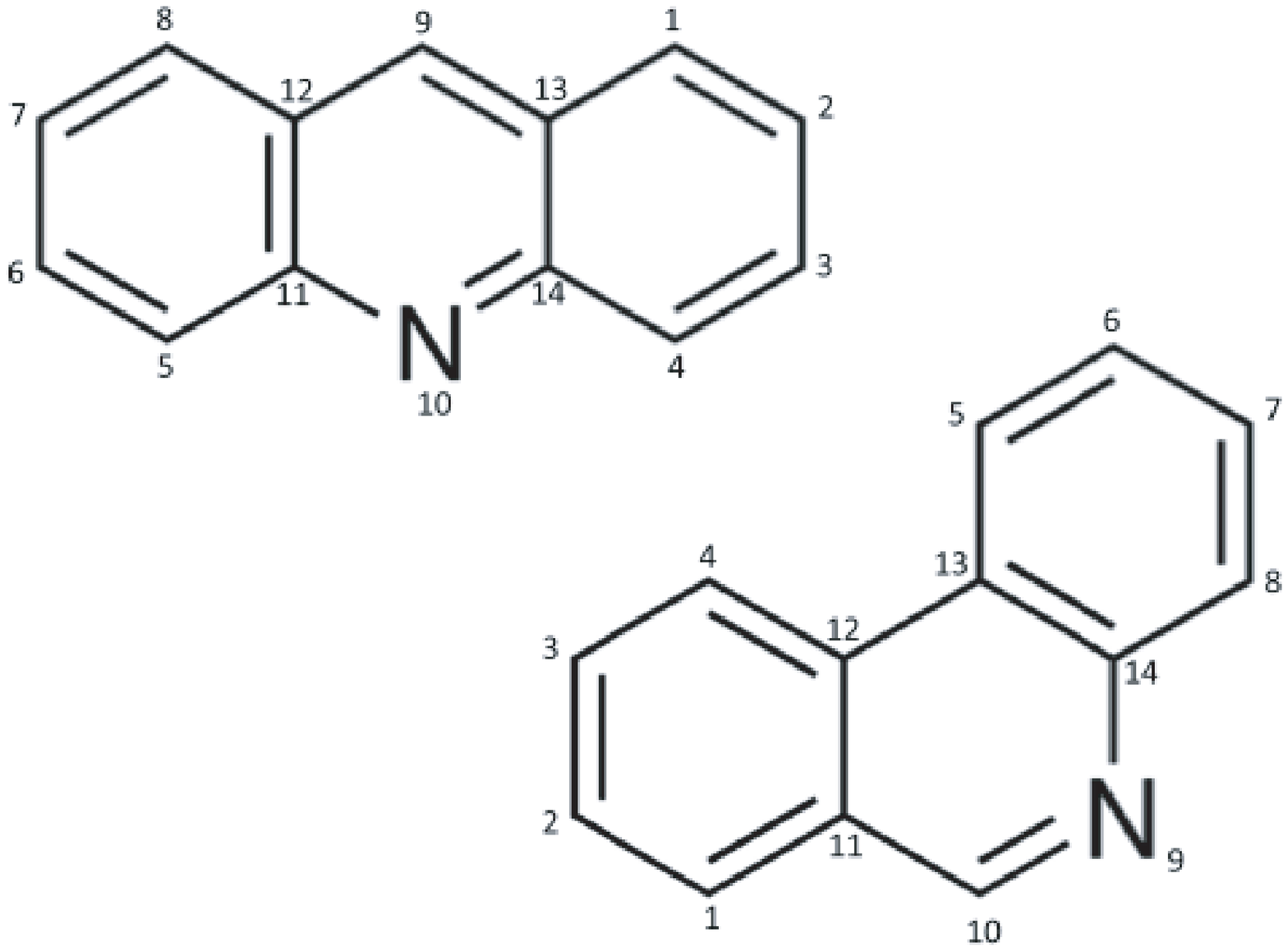}
\caption{Numbering of atoms in acridine and phenanthridine used in this work (note that in these molecules, positions 9 and 10 are equivalent).}\label{fig:structures}
\end{figure*}

\begin{table*}
\caption{AcridineH$^+$ optimised geometry for the ground (S$_0$) and first excited (S$_1$) states, given in Cartesian coordinates. The columns are: the atom type; the x coordinate; the y coordinate; the z coordinate. All coordinates are in Angstroms. To visualise the geometry of the molecule, copy this table into quantum chemistry visualisation software (e.g. molden, molekel, gausview, gabedit, molview.)}
\begin{tabular}{cccc}
\hline\hline
S$_0$ & & &\\
\hline
C &  -.15633932804028E-01 & -.10862080610293E-03 & 0.12980651323101E-01\\
C &  0.21604311219701E-02 & -.19089158226843E-03 & 0.14432770068046E+01\\
C &  0.11972021055134E+01 & 0.89803276031878E-05 & 0.21430993948263E+01\\
C &  0.24199216753256E+01 &0.23481628077462E-03 & 0.14169482086920E+01\\
C &  0.24217097002961E+01 & 0.21387517383260E-03 &  -.22952726592123E-01\\
C &  0.11637565441991E+01 & 0.63721063504643E-04 & -.70947996359481E+00\\
C &  0.36588827993023E+01 & 0.89670036868298E-04 & -.69413240849277E+00\\
C &  0.48768607333507E+01 & 0.99671491006516E-04 & 0.11281256027584E-01\\
C &  0.48385023831464E+01 & 0.11557640267375E-03 & 0.14506720773022E+01\\
N &  0.36204394100487E+01 & 0.26148585104960E-03 & 0.20629420747075E+01\\
C &  0.61534664832050E+01 & 0.39462739087322E-04 & -.63990465747766E+00\\
C &  0.73122559166118E+01 & -.85881352283063E-04 & 0.11515796681357E+00\\
C &  0.60405002971524E+01 & -.34451895911368E-04 & 0.22106321594476E+01\\
H &  0.12086979465670E+01 & 0.27318489630465E-05 & 0.32385711719424E+01\\
H &  -.94338837841307E+00 & -.43462856205420E-03 & 0.19943046581554E+01\\
H &  -.97644198045926E+00 & -.12037658467841E-03 & -.51018374593142E+00\\
H &  0.11594710950112E+01 & 0.92812616426702E-04 & -.18039781566228E+01\\
H &  0.36741793614736E+01 & -.14223620255338E-03 & -.17912511773748E+01\\
H &  0.59984611522114E+01 & -.24205575965616E-04 & 0.33053573920256E+01\\
H &  0.82872752117088E+01 & -.10482312704436E-03 & -.38101766278469E+00\\
C &  0.72545895882594E+01 & -.16468849998998E-03 &  0.15444017023691E+01\\
H &  0.61882703893466E+01 & 0.12746708578654E-03 & -.17338576901536E+01\\
H &  0.81844057585583E+01 & -.36769430733588E-03 & 0.21215810031686E+01\\
H &  0.36061730178587E+01 & 0.47822758248642E-03 & 0.30861488381914E+01\\
\hline
S$_1$ & & &\\
\hline
C &  -.70929314279567E-01 & -.31541786688268E-03 &  0.44608275154867E-01\\
C &  -.58897373717539E-01 & -.13395713442835E-03 & 0.14283873520112E+01\\
C &  0.11957023956751E+01 & 0.15274479310515E-03 & 0.21215288821553E+01\\
C &  0.23990322283165E+01 & 0.34800353268296E-03 & 0.14062986942649E+01\\
C &  0.24044599532987E+01 & 0.40680857123459E-03 & -.25754771470108E-01\\
C &  0.11674457458772E+01 & 0.24263938559302E-03 & -.68146830848166E+00\\
C &  0.36594857191751E+01 & 0.15905476910033E-03 & -.73732905055092E+00\\
C &  0.48941828905466E+01 & -.10639780193868E-03 & 0.89621262121113E-02\\
C &  0.48596781329065E+01 & -.13721149479062E-03 & 0.14406105782956E+01\\
N &  0.36202263549778E+01 & 0.86955555872995E-04 & 0.20781472817843E+01\\
C &  0.61489971947025E+01 & -.10031790764995E-03 & -.61200844532944E+00\\
C &  0.73666490567941E+01 & -.12154090264001E-03 & 0.14831140098671E+00\\
C &  0.60425997832978E+01 & -.10656543469711E-03 & 0.21891108230731E+01\\
H &  0.12096382514204E+01 & 0.25144041827081E-03 & 0.32181100284208E+01\\
H &  -.98921241395055E+00 & -.43914071065955E-04 & 0.20038776168137E+01\\
H &  -.10140993189400E+01 & -.11266848263259E-02 & -.50972845019095E+00\\
H &  0.11482133821656E+01 & 0.76182120393842E-03 & -.17777117058225E+01\\
H &  0.36747793130142E+01 & 0.15942119712985E-03 & -.18340692796154E+01\\
H &  0.59980989482313E+01 & -.11221401574502E-03 & 0.32848772163235E+01\\
H &  0.83249074127402E+01 & -.30888517421707E-03 & -.37951423721326E+00\\
C &  0.73160369744989E+01 & -.36188146590735E-04 & 0.15312167074419E+01\\
H &  0.61987873407062E+01 & -.70651868280096E-04 & -.17072894052606E+01\\
H &  0.82299457308381E+01 & 0.15220479748852E-03 & 0.21324210156096E+01\\
H &  0.36059893202970E+01 & 0.48852424653001E-04 & 0.30990030281594E+01\\

\hline
\end{tabular}
\end{table*}

\begin{table*}
\caption{PhenanthridineH$^+$ optimised geometry for the ground (S$_0$) and first excited (S$_1$) states, given in Cartesian coordinates. The columns are: the atom type; the x coordinate; the y coordinate; the z coordinate. All coordinates are in Angstroms. To visualise the geometry of the molecule, copy this table into quantum chemistry visualisation software (e.g. molden, molekel, gausview, gabedit, molview.)}
\begin{tabular}{cccc}
\hline\hline
S$_0$ & & &\\
\hline
C &  -.29111195181980E+01 & -.65146489862514E+00 & -.61844760890911E-04\\
C &  -.29387908722540E+01 & -.20410167460978E+01 & -.82569277299353E-05\\
C &  -.16588107991563E+01 & 0.77733855689781E-02 &  0.75458641293230E-04\\
C &  -.17219813446285E+01 & -.27709323082638E+01 & -.37199921828215E-04\\
C &  -.41886029648147E+00 & -.69987309161869E+00 & 0.10899941288753E-03\\
C &  -.49024121706251E+00 & -.21199493808559E+01 & -.99735744507177E-05\\
C &  -.51015052010200E+00 & 0.21191676941569E+01 & 0.23906139786150E-04\\
C &  0.75985855123720E+00 & 0.14869008357507E+01 & 0.45563123431349E-04\\
C &  0.82224601841610E+00 &  0.51914718321358E-01 & 0.86731052149833E-04\\
H &  -.38388671239447E+01 & -.68177206849797E-01 & -.29742467374202E-03\\
H &  -.38969462560363E+01 & -.25683393257787E+01 & 0.15197117974559E-03\\
H &  -.17502597060224E+01 & -.38647712115249E+01 & -.50293301186261E-04\\
H &  0.42725694029513E+00 & -.27127154355780E+01 & -.12097218655790E-03\\
H &  -.62271962043965E+00 & 0.32076991915796E+01 & -.14862031904349E-04\\
C &  0.19418691452044E+01 & 0.22795093636996E+01 & 0.24140782153817E-04\\
C &  0.21063891194358E+01 & -.55155814726827E+00 & 0.95493200063119E-04\\
C &  0.32588091784574E+01 & 0.23370986681866E+00 & -.13841075563305E-04\\
C &  0.31814283155797E+01 & 0.16504812920342E+01 & -.66284754156874E-04\\
H &  0.18625496555808E+01 & 0.33716789555816E+01 & 0.11820081928348E-03\\
H &  0.22046994851041E+01 & -.16397107772172E+01 & 0.21478512278879E-03\\
H &  0.42392124513326E+01 & -.25305100290515E+00 & -.93056156862188E-04\\
H &  0.40988144559975E+01 & 0.22465926464354E+01 & -.22937380057018E-03\\
N &  -.16236142038085E+01 & 0.13967651401319E+01 & 0.41276847684163E-04\\
H &  -.25207748385068E+01 & 0.18893664425044E+01 & 0.19856844428137E-04\\
\hline
S$_1$ & & &\\
\hline
C &  -.28919722947996E+01 & -.65278718807287E+00 & 0.33414688738087E-04\\
C &  -.29004845031867E+01 & -.20743437122434E+01 & -.25288045960361E-04\\
C &  -.16856282174091E+01 & 0.51021196541972E-01 & 0.93990452289923E-04\\
C &  -.16975565658929E+01 & -.27971634118656E+01 & -.51094140371324E-04\\
C &  -.43113733728882E+00 & -.66970374682184E+00 & 0.94905579705874E-04\\
C &  -.48109651644802E+00 & -.20981763650497E+01 & 0.10747059410705E-04\\
C &  -.53502759597706E+00 & 0.21924497474230E+01 & -.20002232469464E-04\\
C &  0.73493449624166E+00 & 0.15208906115057E+01 & -.14872482318922E-04\\
C &  0.80956248952303E+00 & 0.80419937319906E-01 & 0.43867431595168E-04\\
H &  -.38417639939713E+01 & -.10589960320023E+00 & 0.24604203282430E-04\\
H &  -.38625961765093E+01 & -.25968695105859E+01 & -.21430470646605E-04\\
H &  -.17040080213666E+01 & -.38902513999984E+01 & -.11837063774503E-03\\
H &  0.45192504967185E+00 & -.26674086478333E+01 & -.21868947529171E-04\\
H &  -.63965656567796E+00 & 0.32799929839071E+01 & -.66438530599537E-04\\
C &  0.19303418180185E+01 & 0.22476705272306E+01 & -.50108855842193E-04\\
C &  0.20903107675528E+01 & -.54987302878548E+00 & 0.37158316542311E-04\\
C &  0.32945230481049E+01 & 0.19861737786567E+00 & 0.20321455353585E-04\\
C &  0.32072629909813E+01 & 0.15900369029004E+01 & -.10390058308755E-04\\
H &  0.18968154620148E+01 & 0.33434241873007E+01 & -.11722732237866E-03\\
H &  0.21566285807046E+01 & -.16413683816682E+01 & 0.23874601091976E-04\\
H &  0.42613150554966E+01 & -.31115225999295E+00 & 0.33851210109629E-04\\
H &  0.41127160537734E+01 & 0.22058439259129E+01 & 0.20611787202523E-04\\
N &  -.16871121931657E+01 & 0.14344031913899E+01 & 0.37651877056406E-04\\
H &  -.25882968303904E+01 & 0.19102256668200E+01 & 0.42093061754362E-04\\
\hline
\end{tabular}
\end{table*}

\clearpage

\section{Mass spectra}\label{mass_spectra}

\begin{figure*}[!htb]
\includegraphics[width=0.9\textwidth]{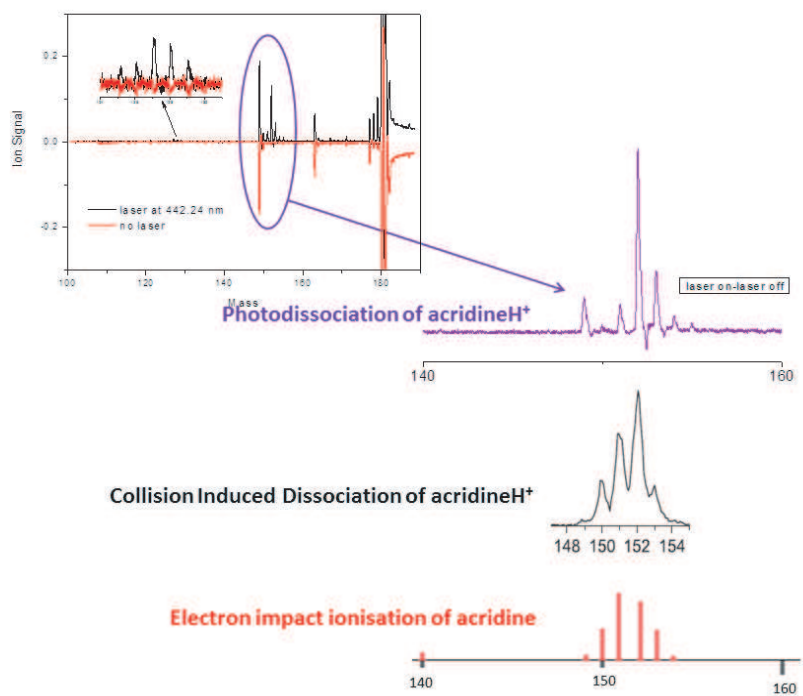}
\caption{AcridineH$^+$ photofragmentation mass spectrum and comparison with collision induced fragmentation and electron impact spectrum of neutral acridine. The first panel shows the experimental mass spectra: the mass spectrum without fragmentation laser is shown in red (and is multiplied by -1 to facilitate comparison), and the mass spectrum recorded with the fragmentation laser on is shown in black. The black spectrum includes the newly formed fragments around m/z 152 (major fragment) and around m/z 127 (minor fragments). The violet curve presents a zoom of the region around m/z 152, presented as a difference spectrum of the recorded spectra with laser on - laser off. This photofragmentation spectrum is different from the mass spectrum obtained by collision induced dissociation of AcH$^+$ (plotted in black), in which the fragments m/z 150 and 151 have much higher intensity. It also differs from the mass spectrum obtained by electron impact ionisation of neutral acridine (plotted in red) which peaks at m/z 151.}\label{fig:tof}
\end{figure*}

\begin{figure*}[!htb]
\includegraphics[width=0.5\textwidth]{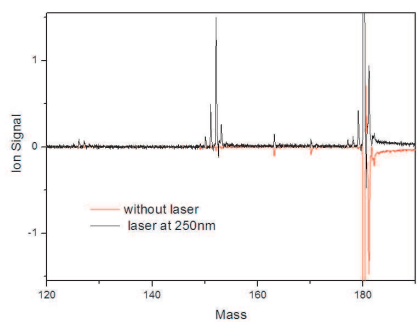}
\caption{PhenanthridineH$^+$ photofragmentation mass spectrum: the red trace corresponds to the mass spectrum recorded without the fragmentation laser, the black trace is the mass spectrum recorded with the fragmentation laser on and shows the appearance of fragments around m/z 152 (major fragments) and around m/z 127 (minor fragments). One spectrum is positive and the other negative just to present clearly the difference between the two spectra.}\label{fig:tofPhH+}
\end{figure*}

\clearpage

\section{PhenanthridineH$^+$ vibronic bands}\label{PhH+vibronic}

\begin{table*}
\caption{PhenanthridineH$^+$ vibronic bands: only the bands with large intensities are tabulated. Above 26500~cm$^{-1}$, the assignments are only tentative.}
\begin{tabular}{ccccc}
\hline\hline
Wavelength (nm) &	Wavenumbers (cm$^{-1}$) &	Energy above origin (cm$^{-1}$) &	Intensity &	Assignment\\
\hline
397.28 &	25171 &	0	  & m   &	0$_0^0$\\
393.48 &	25414 &	243	  & m   &	5$_0^1$\\
389.8  &	25654 &	483	  & m   &	5$_0^2$ \& 15$_0^1$\\
389.04 &	25704 &	533	  & str  &	17$_0^1$\\
386.16 &	25896 &	725	  & m   &	5$_0^3$ \& 15$_0^1$5$_0^1$\\
385.44 &	25944 &	773	  & str &	5$_0^1$17$_0^1$\\
382.68 &	26131 &	960	  & m  &	5$_0^4$\\
381.84 &	26189 &	1018  &	str &	5$_0^2$17$_0^1$\\
381.16 &	26236 &	1065  &	str &	17$_0^2$\\
379.64 &	26341 &	1170  &	m &	\\
378.36 &	26430 &	1259  &	m &	5$_0^3$17$_0^1$\\
377.68 &	26477 &	1306  &	str &	5$_0^1$17$_0^2$\\
376.72 &	26545 &	1374  &	str &	\\
376.2  &	26582 &	1411  &	str &	\\
374.92 &	26672 &	1501  &	m &	\\
374.28 &	26718 &	1547  &	str & 	5$_0^2$17$_0^2$\\
373.56 &	26769 &	1598  &	str &	17$_0^3$\\
372.12 &	26873 &	1702  &	str &	\\
371.68 &	26905 &	1734  &	m &	\\
370.88 &	26963 &	1792  &	m &	\\
369.88 &	27036 &	1865  &	str &	\\
369.4 &	27071 &	1900  &	str &	\\
368.84 &	27112 &	1941  &	str &	\\
367.64 &	27201 &	2030  &	m &	\\
366.2 &	27307 &	2136  &	str &	17$_0^4$\\
362.24 &	27606 &	2435  &	str &	\\
\hline
\end{tabular}
\end{table*}

\end{document}